\documentclass[usenatbib,useAMS]{mn2e}
\usepackage{natbib,times}
\usepackage{longtable}
\usepackage{amssymb}
\usepackage{rotfloat}
\usepackage{rotating}
\usepackage{epsfig}
\usepackage{booktabs}
\usepackage{amsmath}
\usepackage{graphicx}
\usepackage{makecell}
\def\hMpc{~h^{-1}{\rm Mpc}}
\def\kms{km~s$^{-1}$}
\def\HI{H\textsc{i} }

\begin{document}

\title[2MTF - VII. Final data release]
{2MTF - VII. 2MASS Tully-Fisher survey final data release: distances for 2,062 nearby spiral galaxies}

\author[T. Hong et al.]
{Tao~Hong~$^{1,2,3,4}$\thanks{E-mail: hongtao@nao.cas.cn}, 
Lister~Staveley-Smith~$^{3,4,5}$, 
Karen~L.~Masters~$^{6}$, 
\newauthor
Christopher~M.~Springob~$^{3,4,7}$, 
Lucas~M.~Macri~$^{8}$, 
B\"arbel~S.~Koribalski~$^{9}$, 
D.~Heath~Jones~$^{10}$,
\newauthor
Tom~H.~Jarrett~$^{11}$, 
Aidan~C.~Crook~$^{12}$, 
Cullan Howlett~$^{3,4}$~and Fei Qin~$^{3,4}$
\\
$^{1}$National Astronomical Observatories, Chinese Academy
  of Sciences, 20A Datun Road, Chaoyang District, Beijing 100012,
  China\\
$^{2}$CAS Key Laboratory of FAST, National Astronomical Observatories, Chinese Academy of Sciences\\
$^{3}$International Centre for Radio Astronomy Research,
  M468, University of Western Australia, Crawley, 35 Stirling Highway, WA 6009, Australia\\
$^{4}$ARC Centre of Excellence for All-sky Astrophysics
  (CAASTRO)\\
$^{5}$ARC Centre of Excellence for All Sky Astrophysics in 3 Dimensions (ASTRO 3D)\\
$^{6}$Haverford College, Department of Physics and Astronomy, 370 Lancaster Avenue, 
Haverford, Pennsylvania 19041, USA\\
$^{7}$Australian Astronomical Observatory, PO Box 915, North Ryde, NSW 1670 Australia\\
$^{8}$George P. and Cynthia Woods Mitchell Institute for Fundamental Physics and 
Astronomy, Department of Physics and Astronomy, Texas A\&M University, \\4242 TAMU, 
College Station, TX 77843, USA\\ 
$^{9}$CSIRO Astronomy \& Space Science, Australia Telescope National 
Facility, PO Box 76, Epping, NSW 1710, Australia\\ 
$^{10}$English Language and Foundation Studies Centre, University of Newcastle, Callaghan NSW 2308, Australia\\ 
$^{11}$Astronomy Department, University of Cape Town, Private Bag X3. 
Rondebosch 7701, Republic of South Africa\\
$^{12}$Microsoft Corporation, 1 Microsoft Way, Redmond, WA 98052, USA\\
}

\date{Accepted  ... Received  ...}

\pagerange{\pageref{firstpage}--\pageref{lastpage}} \pubyear{}

\maketitle
\label{firstpage}
\begin{abstract}
We present the final distance measurements for the 2MASS Tully-Fisher (2MTF) survey. The final 2MTF 
catalogue contains 2,062 nearby spiral galaxies in the CMB frame velocity range of 600 \kms$< cz < 10,000$ \kms\ 
with a mean velocity of 4,805 \kms. The main update in this release is the replacement of some archival HI data with newer ALFALFA data. Using the 2MTF template relation, we calculate the distances and 
peculiar velocities of all 2MTF galaxies. The mean uncertainties of the linear distance measurements are around 22\% in all three infrared 
bands. 2MTF measurements agree well with the distances from the Cosmicflows-3 compilation, which contains 1,117 common galaxies, including 28 with SNIa distance measurements. 
Using distances estimated from the `3-bands combined' 2MTF sample and a 
$\chi^2$ minimization method, we find best-fit bulk flow amplitudes of $308 \pm 26$ \kms, $318 \pm 29$ \kms, and $286 \pm 25$ \kms at 
depths of $R_I = $ 20, 30 and 40$\hMpc$, respectively, which is consistent with the $\Lambda$CDM model and with previous 2MTF results with different estimation techniques and a preliminary catalogue.
\end{abstract}

\section{Introduction}
The matter distribution in the Universe appears to be homogeneous and isotropic on large scales (a.k.a. the cosmological principle). However, on relatively small scales, structures such as clusters, filaments, walls and voids dominate the matter distribution \citep{Geller1989, Gott2005}. 
Studies of cosmological structures or lack thereof provide tight constraints on the cosmological model 
\citep{Bernardeau2002, Jones2009,Scrimgeour2012,Lombriser2012, Alpaslan2014}. Useful information 
on the matter distribution field can be obtained from galaxy redshift surveys, gravitational lensing studies and X-ray and Sunyaev-Zel'dovich cluster searches \citep[e.g.][and references therein]{Eisenstein2005, Hong2016, 
Bautista2017, Waerbeke2013, Planck2014,Salvati2018}.
The peculiar motions of galaxies induced by inhomogeneous matter distribution are also useful probes for tracing structures. These are typically measured using redshift-independent distances \citep{Springob2007,Masters2008,Campbell2014,Tully2016}, proper motions \citep{Brunthaler2005} or the kinetic Sunyaez-Zel'dovich effect \citep{Kashlinsky2011}. Moreover, such measurements trace both the visible and the dark matter distributions \citep{Hong2014, Scrimgeour2016, Springob2016, Qin2018}.

The line-of-sight peculiar velocity is the non-Hubble component of a galaxy's motion, which can be estimated in the nearby Universe by 
\begin{equation}
v_{pec} = cz - H_0r,
\label{eq:vpec}
\end{equation}
where $H_0$ is the Hubble constant, $z$ is the redshift of the galaxy and $r$ is the redshift independent 
distance. In the context of this paper the Tully-Fisher (TF) relation \citep{Tully1977}, which combines measurements of galaxy luminosity and rotational velocity, provides a good tool to accurately measure redshift independent distances for large numbers of 
spiral galaxies. However, many other distance-determination techniques exist, including supernova type Ia \citep{Phillips1993}, 
Fundamental Plane \citep{Dressler1987} and surface brightness fluctuations \citep{Tonry1988}.

The SFI++ catalog \citep{Springob2007} is the largest TF sample to date, and includes peculiar velocity information 
for 4,861 spirals. Because of the effect of extinction in its input catalogues, SFI++ contains 
a large gap close to the Galactic plane ($|b|<15^\circ$). It therefore misses significant parts of massive nearby structures such as the Hydra-Centaurus, Norma and Vela superclusters and the Great Attractor \citep{Kraan-Korteweg2000,Kraan-Korteweg2017}, whilst introducing possible biases in cosmological parameters derived
from peculiar velocity fields measured in a non-spherical region \citep{Andersen2016}. 

The 2MASS Tully-Fisher Survey \citep[2MTF,][]{Masters2008, Hong2013, Masters2014} uses the 
infrared magnitudes in the $J$, $H$ and $K$ bands from the 2MASS Redshift survey \citep[2MRS,][]{Huchra2012} in order
to reduce the extinction effect due to the Galactic plane, leaving only a small Zone of Avoidance at Galactic 
latitudes of $|b| \leq 5^\circ$. Combined with high-quality 21-cm \HI spectral data, the final 2MTF sample provides 
2,062 peculiar velocity measurements for nearby spirals. In this paper, we present the final 2MTF sample and describe the 2MTF data collection and reduction process in Section~\ref{sec:collect}. The final catalog 
is presented in Section~\ref{sec:catalog}. 
Comparisons between 2MTF distances and Cosmicflows-3 measurements are 
made in Section~\ref{sec:compare}. Finally, we update our previous calculations of the bulk flow amplitude for the 2MTF sample in Section~\ref{sec:bulk}.

\section{Observational data and methodology}
\label{sec:collect}
The 2MTF survey is a Tully-Fisher survey of nearby spiral galaxies, producing high-quality 
redshift independent distance and peculiar velocity measurements. The observational methods and data reduction are described in detail by \citet{Hong2014}, which we briefly summarise below.

\subsection{Photometric data}
All target galaxies in 2MTF are selected from the 2MRS catalog with the following criteria: total $K$-band magnitude $K < 11.25$ mag, velocity $cz < 10,000$ \kms, and 2MASS co-added 
axis ratio $b/a < 0.5$. About 6,600 2MRS galaxies meet the selection criteria and became our
2MTF target galaxies. Because the TF template relations vary with the type of spiral galaxy \citep{Masters2008}, we also adopt the morphological type code $T$ to calibrate the galaxies to the correct template relations. The co-added axis ratio used above formally differs from 
the $I$-band axis ratio used by \citet{Masters2008} when 
building the TF template relation. This may introduce small differences in the final measurements, 
but with no significant bias \citep{Hong2014}. We follow \citet{Masters2008} 
in making galaxy internal dust extinction and $k$-corrections.

\subsection{{\rm H}\,{\sc i} spectra widths}
High-quality galaxy rotation widths as measured from \HI spectra are a key part of accurate TF measurements. The 2MTF \HI spectral data derive from our own new observations 
using the GBT and Parkes telescopes, new observations provided by the ALFALFA survey, and high signal-to-noise ratio archival data. 

\subsubsection{GBT and Parkes observations}
The 2MTF project observed 1,193 target galaxies in the sky area of $\delta > -40^{\circ}$ using the 
Green Bank Telescope using a position-switching mode \citep{Masters2014}. 727 galaxies were detected 
in H\textsc{i}, and 383 of them are of a quality that allows them to be included into the final 2MTF catalogue. All the `well-detected' galaxies have signal-to-noise ratio S/N $\geq 10$, and show normal \HI profiles consistent with non-interacting, non-confused spiral galaxies. In addition, all galaxies meet further data quality selection criteria as described in Section~\ref{sec:quality}. The smoothed 
GBT velocity resolution is 5.15~\kms.

For $\delta \leq -40^{\circ}$, the Parkes radio telescope was employed to conduct  \HI observations, also in position-switching mode, but one in which the target galaxy was always observed by one of seven central beams of the Parkes multibeam receiver \citep{Hong2014}. 
From the 305 target galaxies, we obtained 110 high signal-to-noise ratio \HI spectra which were included 
into the final 2MTF sample. The Parkes \HI spectra provide 
a velocity resolution of $3.3$~\kms~after Hanning smoothing. 

All \HI spectra obtained by GBT and Parkes telescope were measured using the IDL routine \textit{awv\_fit.pro}. This 
routine measures \HI line widths using several methods. The W$_{F50}$ width is adopted as our preferred width for this study.

\subsubsection{ALFALFA data}
Duplication of observations in the northern hemisphere was avoided by leaving the 7000 deg$^2$ covered in the ALFALFA survey \citep{Giovanelli2005}. Overall, ALFALFA contains 
$\sim 31,500$ extragalactic \HI sources with redshift out to $z < 0.06$ \citep{Haynes2018}. We extracted 545 high-quality spectra and \HI widths for galaxies which passed the 2MTF selection criteria from the complete ALFALFA catalog. For consistency with the GBT and Parkes analysis, we used the W$_{F50}$ width measurements.

\subsubsection{Archival data}
Besides our own observations and the ALFALFA data, we also used \HI widths from archival sources where the data was high quality. The largest archival data source we adopted was the Cornell \HI digital archive \citep{Springob2005} of \HI observations of 9,000 galaxies  ($cz < 28,000$ \kms), obtained from single-dish 
telescopes. According to the \HI line quality, \citet{Springob2005} marked the \HI lines with codes 
G (Good), F (Fair), S (Single peak) and P (Poor). Only G (Good) and F (Fair) galaxies were accepted into the 2MTF catalog. A total of
711 \HI widths from the Cornell \HI digital archive were included. 

\HI width measurements from other sources were also included \citep{Theureau1998, Theureau2005, Theureau2007, Mathewson1992, Paturel2003}. We collected raw \HI widths from these catalogs and corrected for observational 
effects following the same procedure as for our own observations \citep[for more details, see][]{Hong2013}.

\subsubsection{Data Selection}
\label{sec:quality}
To improve the distance measurement accuracy, further data quality limits were adopted. Galaxies with $cz < 600$ \kms\ (CMB frame), 
\HI spectra with S/N ratio $ < 5$, or relative \HI error 
$\epsilon_\mathrm{w}/w_{\mathrm{HI}} > 10\%$ were omitted. 
When one galaxy was observed with more than one telescope, we preferred the newer data, i.e., our own observations and 
ALFALFA data have higher priority than the archival data. 

2MASX J00383973+1724113 has a reported 2MASS   
axis ratio of $b/a$ = 0.36, which doesn't represent the 
outer blue disk present in the galaxy. We give this galaxy 
a flag of `W' in final data table, for optional exclusion.

\subsection{Tully-Fisher distances}
For calculating the Tully-Fisher distances and peculiar velocities for 2MTF galaxies, we use the TF template relations developed by \citet{Masters2008}:
\begin{equation}
\begin{aligned}
M_K - 5\log h=-22.188-10.74(\log W -2.5),\\
M_H - 5\log h=-21.951-10.65(\log W -2.5),\\
M_J - 5\log h=-21.370-10.61(\log W -2.5),
\end{aligned}
\label{eq:tf}
\end{equation}
where  $M_K$, $M_H$, and $M_J$ are the absolute magnitudes and $W$ is the corrected \HI linewidth. The TF template relation depends on the galaxy morphology in all bands; the above relations refer to galaxies of type Sc. 

As described in \citet{Hong2014}, the uncertainties of TF peculiar velocities are log-normal. So the preferred way to quote distances for 2MTF is  
the logarithmic quantity $\log(d_{z}/d_{\mathrm{TF}})$, 
\begin{equation}
\log\left(\frac{d_z}{d^{*}_{\mathrm{TF}}}\right)=\frac{-\Delta M}{5},
\label{eq:dm}
\end{equation}
where $d_{z}$ is the redshift distance of the galaxy in the CMB frame, $d^{*}_{\mathrm{TF}}$ is the galaxy's redshift-independent distance from the TF relation before Malmquist bias correction (which will be corrected in the final step of the calculation), and
$\Delta M = M_{obs}-M(W)$ is the difference between the absolute magnitude calculated using a redshift distance and the absolute magnitude predicted by the TF template relations of Equation~\ref{eq:tf}. The corrected absolute 
magnitude $M_{obs}$ contains the terms for $K$-correction ($k_X$), 
extinction correction due to the Galaxy ($A_X$), internal 
extinction correction of the galaxy itself ($I_X$) and the morphological type correction $T_X$ which arises from the different 
slopes and zero points of the TF relation between different types of galaxies, we adopt the same equation as \citet{Masters2008} of 
\begin{equation}
M_{obs} - 5\log h = m_{obs} + k_X -A_X-I_X-T_X-5\log v_{CMB}-15,
\end{equation}
where $X$ represents $K, H$ or $J$, $v_{CMB}$ is the recession velocity of the galaxy in the CMB frame. We make all corrections following 
\citet{Masters2008}.

The error in this logarithmic quantity includes four components: intrinsic error, inclination error, \HI width error and NIR magnitude error \citep[see Section 3.2 and Appendix A of][for more details]{Hong2014}, 
where we adopt an intrinsic error inversely proportional to the HI width to represent 
the fact that the scatter of the TF relation decreases with the \HI width \citep{Giovanelli1997, Masters2006}. 
The error in redshift is considered negligible because of its much higher accuracy. The final error in the 
logarithmic quantity is the sum in quadrature of all components in logarithmic units. To improve measurement accuracy, we used the galaxy group information presented by \citet{Crook2007} to assign the same redshift distance $d_{z}$ to all galaxies in the same group, whilst $\log(d_{z}/d_{\mathrm{TF}})$ was calculated separately.

The final step of data reduction is the Malmquist bias correction. There are two types of Malmquist bias. Homogeneous Malmquist bias which arises from distance-dependent selection effects. This bias affects all galaxies regardless of their position. Inhomogeneous Malmquist bias is caused by the variation of density along the line of sight. As described by \citet{strauss1995}, this bias is smaller in redshift space than distance space because of much smaller measurement errors. Thus we corrected for homogeneous Malmquist bias only, and considered the inhomogeneous bias to be negligible.
We divided the sample into two parts north and south of declination $\delta = -40^\circ$, which reflects the fact that the completeness is different in these two sky regions. Corrections were made using the procedure of \citet{Hong2014}, which in turn follows the procedure of \citet{Springob2014}. 

\section{Catalog presentation}
\label{sec:catalog}
We present the photometric data from 2MRS catalog and corrected \HI widths for the 2,062 2MTF galaxies in Table~\ref{tab:data}. The definition of the columns of the table is as follows:\\

Column (1). --- The 2MASS XSC ID name.

Column (2) and (3). --- Right ascension (RA) and declination (DEC) in
the J2000.0 epoch from the 2MASS XSC.

Column (4). --- The barycentric redshift $cz_{\rm 2MRS}$ from the 2MRS (\kms).

Column (5 - 7). --- The NIR magnitudes in the $K$, $H$ and $J$ bands from the 2MASS XSC, respectively.

Column (8 - 10). --- The errors of the NIR magnitudes in $K$, $H$ and $J$ bands from the 2MASS XSC.

Column (11). --- The corrected \HI width (\kms).

Column (12). --- The error of corrected \HI width (\kms).\\

\begin{table*}
\caption[]{Photometric and corrected \HI width data for 2MTF galaxies.}
\label{tab:data}
\centering
\begin{tabular}{crrrrrrrrrrr}
\hline
\hline
2MASX-ID	&	 \multicolumn{1}{c}{RA}	   &    \multicolumn{1}{c}{DEC}	& $cz_{2MRS}$   & $m_K$   &  $m_H$   &  $m_J$	& $\epsilon_{m_K}$  & $\epsilon_{m_H}$  & $\epsilon_{m_J}$  &   $W_c$  & $\epsilon_{W_c}$\\
	&	 \multicolumn{1}{c}{[deg] (J2000)} 	   &   \multicolumn{1}{c}{[deg] (J2000)} &  [\kms]   & [mag]   &  [mag]   &  [mag]	& [mag]  & [mag]  & [mag]  &   [\kms]  & [\kms]\\
\hline
00005604+2020165&$     0.23345$&$    20.33799$&$    6804$&$  11.124$&$  11.616$&$  12.339$&$ 0.061$&$ 0.062$&$ 0.046$&$   406$&$   5.1$\\
00005891+2854421&$     0.24550$&$    28.91172$&$    6899$&$  10.818$&$  11.085$&$  11.515$&$ 0.059$&$ 0.048$&$ 0.029$&$   440$&$  11.8$\\
00010478+0432261&$     0.26987$&$     4.54060$&$    9151$&$  11.029$&$  11.274$&$  11.919$&$ 0.047$&$ 0.036$&$ 0.026$&$   476$&$  23.0$\\
00011976+3431326&$     0.33238$&$    34.52571$&$    5032$&$  10.405$&$  10.704$&$  11.384$&$ 0.037$&$ 0.028$&$ 0.023$&$   439$&$   5.9$\\
00013830+2329011&$     0.40968$&$    23.48363$&$    4371$&$   9.330$&$   9.591$&$  10.327$&$ 0.024$&$ 0.021$&$ 0.019$&$   544$&$   8.0$\\
00014193+2329452&$     0.42471$&$    23.49588$&$    4336$&$   9.930$&$  10.154$&$  11.056$&$ 0.025$&$ 0.021$&$ 0.019$&$   635$&$  10.6$\\
00024636+1853106&$     0.69321$&$    18.88622$&$    7882$&$  10.776$&$  10.935$&$  11.652$&$ 0.042$&$ 0.032$&$ 0.027$&$   518$&$  11.2$\\
00035889+2045084&$     0.99545$&$    20.75235$&$    2309$&$   8.400$&$   8.700$&$   9.438$&$ 0.008$&$ 0.007$&$ 0.006$&$   384$&$   6.0$\\
00041299+1047258&$     1.05410$&$    10.79052$&$    7887$&$  10.678$&$  10.943$&$  11.652$&$ 0.045$&$ 0.036$&$ 0.029$&$   412$&$  12.2$\\
00051672-1628368&$     1.31967$&$   -16.47691$&$    7412$&$  10.135$&$  10.448$&$  11.174$&$ 0.028$&$ 0.026$&$ 0.020$&$   528$&$  12.1$\\
00055236+2232083&$     1.46812$&$    22.53567$&$    6644$&$  11.221$&$  11.520$&$  12.320$&$ 0.045$&$ 0.039$&$ 0.037$&$   273$&$   6.5$\\
00064016+2609164&$     1.66731$&$    26.15452$&$    7556$&$  10.965$&$  11.338$&$  11.938$&$ 0.045$&$ 0.044$&$ 0.034$&$   393$&$  14.7$\\
00071951+3236334&$     1.83139$&$    32.60925$&$    5076$&$   9.865$&$  10.136$&$  10.914$&$ 0.026$&$ 0.022$&$ 0.019$&$   461$&$   4.8$\\
00080679+0943037&$     2.02831$&$     9.71773$&$    6423$&$  10.523$&$  10.826$&$  11.450$&$ 0.049$&$ 0.041$&$ 0.033$&$   408$&$   9.2$\\
00084772+3326000&$     2.19883$&$    33.43332$&$    4808$&$   9.731$&$   9.974$&$  10.639$&$ 0.027$&$ 0.024$&$ 0.020$&$   457$&$   5.5$\\
00090421+1055081&$     2.26760$&$    10.91890$&$    6674$&$  10.605$&$  10.974$&$  11.870$&$ 0.043$&$ 0.035$&$ 0.033$&$   464$&$   4.6$\\
00092866+4721208&$     2.36937$&$    47.35583$&$    5153$&$  11.218$&$  11.562$&$  12.181$&$ 0.066$&$ 0.062$&$ 0.045$&$   308$&$  11.8$\\
00093281+4808068&$     2.38672$&$    48.13519$&$    5388$&$  11.018$&$  11.229$&$  11.965$&$ 0.046$&$ 0.039$&$ 0.032$&$   384$&$  11.2$\\
00102637+2859164&$     2.60990$&$    28.98794$&$    7850$&$  11.171$&$  11.552$&$  12.460$&$ 0.046$&$ 0.041$&$ 0.039$&$   432$&$   8.2$\\
00103277+2859464&$     2.63654$&$    28.99624$&$    7033$&$  10.409$&$  10.648$&$  11.435$&$ 0.033$&$ 0.024$&$ 0.024$&$   428$&$  24.3$\\
00111259-3334428&$     2.80248$&$   -33.57855$&$    7853$&$  10.295$&$  10.601$&$  11.187$&$ 0.044$&$ 0.034$&$ 0.021$&$   503$&$  41.9$\\
00121573+2219187&$     3.06553$&$    22.32180$&$    7629$&$  10.584$&$  10.896$&$  11.571$&$ 0.034$&$ 0.024$&$ 0.023$&$   514$&$   9.3$\\
00143187-0044156&$     3.63279$&$    -0.73760$&$    3946$&$  11.028$&$  11.312$&$  11.984$&$ 0.069$&$ 0.056$&$ 0.046$&$   310$&$   7.1$\\
00144010+1834551&$     3.66711$&$    18.58203$&$    5392$&$   9.479$&$   9.843$&$  10.654$&$ 0.024$&$ 0.016$&$ 0.015$&$   554$&$   5.5$\\
00155127+1605232&$     3.96365$&$    16.08977$&$    4213$&$  10.509$&$  10.813$&$  11.550$&$ 0.040$&$ 0.028$&$ 0.022$&$   298$&$   6.6$\\
00164417+0704337&$     4.18407$&$     7.07609$&$    3966$&$   9.750$&$  10.099$&$  10.873$&$ 0.022$&$ 0.016$&$ 0.016$&$   387$&$   4.8$\\
00165087-0516060&$     4.21202$&$    -5.26830$&$    3968$&$  10.510$&$  10.800$&$  11.449$&$ 0.055$&$ 0.044$&$ 0.033$&$   268$&$  21.0$\\
00180131-5905008&$     4.50560$&$   -59.08359$&$    8924$&$  11.072$&$  11.464$&$  12.346$&$ 0.053$&$ 0.041$&$ 0.039$&$   443$&$  22.5$\\
00181053+1817323&$     4.54393$&$    18.29227$&$    5521$&$  11.219$&$  11.439$&$  12.253$&$ 0.058$&$ 0.031$&$ 0.034$&$   349$&$   5.1$\\
00183335-0616195&$     4.63899$&$    -6.27206$&$    6520$&$  11.116$&$  11.436$&$  12.205$&$ 0.044$&$ 0.037$&$ 0.033$&$   277$&$  12.0$\\

\hline
\multicolumn{12}{l}{Table~\ref{tab:data} is available in its entirety online. A portion is shown here for guidance regarding its form and content.}
\end{tabular}
\end{table*}

Table~\ref{tab:distance} contains the logarithmic distance quantity measured by 2MTF, together with linear 
peculiar velocities defined by Eq.~(\ref{eq:vpec}), the latter is provided for convenience. 
We also include the peculiar velocities generated from the estimator introduced by 
\citet{Watkins2015}: 
\begin{equation}
V_{WF15} = cz\ln{\left(cz/H_0r\right)}=cz\ln(d_{z}/d_{\mathrm{TF}}), 
\label{eq:wf15}
\end{equation}
where $z$ is the redshift of the galaxy and $r$ is the redshift-independent distance (the Tully-Fisher distance in this paper). 
We adopt a low redshift approximation of the accurate WF15 estimator (Eq. 7 of \citealt{Watkins2015}) here, 
since the redshift of 2MTF galaxies 
is much smaller than unity.
All measurements are made in the CMB reference frame, so we convert the 
recession velocities of galaxies to this frame using a solar motion of $v = 384$~\kms\,towards to 
$(l, b) = (263.99^\circ, 48.26^\circ)$ \citep{Planck2014b}. Table~\ref{tab:distance} is organized as follows:\\

Column (1). --- The 2MASS XSC ID name.

Column (2). --- The CMB frame redshift $cz_{\rm CMB}$ (\kms).

Column (3 - 5). --- The logarithmic quantities and errors measured by $K$, $H$ and $J$ band magnitude, respectively.

Column (6 - 8). --- The peculiar velocities and errors measured by $K$, $H$ and $J$ band magnitude, respectively (\kms).

Column (9 - 11). --- The peculiar velocities and errors calculated by WF15 estimator in $K$, $H$ and $J$ band magnitude, respectively (\kms).

Column (12). --- Data quality warning flag, a flag `W' reminds one to use the measurement of the galaxy carefully.\\
\begin{table*}
\caption{~Distances and peculiar velocities for 2,062 2MTF galaxies.}
\small
\label{tab:distance}
\resizebox{\linewidth}{!}{

\begin{tabular}{lcrrrrrrrrrc}
\hline\hline
\multicolumn{1}{c}{2MASX ID}  &\multicolumn{1}{c}{$cz_{CMB}$} 
& \multicolumn{1}{c}{$\log\frac{d_{z}}{d_{\mathrm{TF,K}}}$} &\multicolumn{1}{c}{$\log\frac{d_{z}}{d_{\mathrm{TF,H}}}$} & \multicolumn{1}{c}{$\log\frac{d_{z}}{d_{\mathrm{TF,J}}}$} &\multicolumn{1}{c}{$V_{pec, K}$} & \multicolumn{1}{c}{$V_{pec, H}$} &\multicolumn{1}{c}{$V_{pec, J}$} & \multicolumn{1}{c}{$V_{WF15,K}$} & \multicolumn{1}{c}{$V_{WF15,H}$} & \multicolumn{1}{c}{$V_{WF15,J}$} & \multicolumn{1}{c}{Flag}\\
\cline{6-11}

\multicolumn{1}{c}{-~-}  &\multicolumn{1}{c}{[\kms]} &\multicolumn{1}{c}{[-~-]} &\multicolumn{1}{c}{[-~-]} &\multicolumn{1}{c}{[-~-]} &\multicolumn{6}{c}{[\kms]} &\multicolumn{1}{c}{[-~-]}\\

\multicolumn{1}{c}{(1)} & \multicolumn{1}{c}{(2)} & \multicolumn{1}{c}{(3)} & \multicolumn{1}{c}{(4)} & \multicolumn{1}{c}{(5)} & \multicolumn{1}{c}{(6)} & \multicolumn{1}{c}{(7)} & \multicolumn{1}{c}{(8)} & \multicolumn{1}{c}{(9)} & \multicolumn{1}{c}{(10)} & \multicolumn{1}{c}{(11)}& \multicolumn{1}{c}{(12)}\\
\hline
00005604+2020165&$    6439$&$ -0.127\pm  0.078$&$ -0.156\pm  0.075$&$ -0.160\pm  0.080$&$ -2165\pm 389$&$ -2756\pm 476$&$ -2843\pm 524$&$ -1889\pm1152$&$ -2316\pm1110$&$ -2377\pm1185$&--\\
00005891+2854421&$    6546$&$ -0.071\pm  0.092$&$ -0.057\pm  0.088$&$ -0.023\pm  0.093$&$ -1145\pm 243$&$  -907\pm 184$&$  -340\pm  73$&$ -1066\pm1388$&$  -859\pm1334$&$  -342\pm1406$&--\\
00010478+0432261&$    8717$&$ -0.061\pm  0.091$&$ -0.042\pm  0.087$&$ -0.041\pm  0.092$&$ -1265\pm 265$&$  -838\pm 170$&$  -823\pm 174$&$ -1226\pm1827$&$  -845\pm1756$&$  -831\pm1851$&--\\
00011976+3431326&$    4694$&$ -0.092\pm  0.083$&$ -0.081\pm  0.079$&$ -0.081\pm  0.085$&$ -1106\pm 211$&$  -964\pm 175$&$  -956\pm 187$&$  -995\pm 896$&$  -879\pm 857$&$  -873\pm 922$&--\\
00013830+2329011&$    3982$&$ -0.218\pm  0.075$&$ -0.200\pm  0.070$&$ -0.212\pm  0.077$&$ -2582\pm 452$&$ -2318\pm 374$&$ -2491\pm 442$&$ -1996\pm 692$&$ -1833\pm 641$&$ -1941\pm 705$&--\\
00014193+2329452&$    4018$&$ -0.385\pm  0.069$&$ -0.352\pm  0.062$&$ -0.382\pm  0.069$&$ -5768\pm 916$&$ -5053\pm 721$&$ -5717\pm 908$&$ -3559\pm 636$&$ -3254\pm 575$&$ -3538\pm 638$&--\\
00024636+1853106&$    7516$&$ -0.069\pm  0.084$&$ -0.027\pm  0.078$&$ -0.035\pm  0.084$&$ -1252\pm 242$&$  -451\pm  81$&$  -605\pm 117$&$ -1187\pm1449$&$  -467\pm1352$&$  -611\pm1454$&--\\
00035889+2045084&$    1946$&$  0.027\pm  0.087$&$  0.038\pm  0.085$&$  0.035\pm  0.091$&$   120\pm  24$&$   164\pm  32$&$   152\pm  32$&$   123\pm 391$&$   170\pm 380$&$   158\pm 407$&--\\
00041299+1047258&$    7601$&$  0.073\pm  0.097$&$  0.083\pm  0.093$&$  0.078\pm  0.098$&$  1192\pm 266$&$  1341\pm 287$&$  1271\pm 287$&$  1271\pm1696$&$  1449\pm1633$&$  1365\pm1720$&--\\
00051672-1628368&$    7086$&$ -0.111\pm  0.076$&$ -0.102\pm  0.070$&$ -0.116\pm  0.077$&$ -2037\pm 357$&$ -1864\pm 305$&$ -2149\pm 381$&$ -1806\pm1243$&$ -1671\pm1150$&$ -1893\pm1263$&--\\
00055236+2232083&$    6237$&$  0.208\pm  0.117$&$  0.203\pm  0.118$&$  0.181\pm  0.120$&$  2369\pm 638$&$  2330\pm 633$&$  2128\pm 588$&$  2983\pm1675$&$  2920\pm1696$&$  2605\pm1728$&--\\
00064016+2609164&$    7203$&$  0.023\pm  0.092$&$  0.016\pm  0.090$&$  0.030\pm  0.095$&$   389\pm  83$&$   274\pm  57$&$   488\pm 107$&$   386\pm1534$&$   265\pm1489$&$   491\pm1571$&--\\
00071951+3236334&$    4608$&$ -0.083\pm  0.076$&$ -0.064\pm  0.072$&$ -0.076\pm  0.079$&$ -1041\pm 182$&$  -792\pm 131$&$  -942\pm 171$&$  -885\pm 810$&$  -677\pm 760$&$  -804\pm 835$&--\\
00080679+0943037&$    6043$&$  0.057\pm  0.100$&$  0.057\pm  0.098$&$  0.063\pm  0.102$&$   749\pm 172$&$   749\pm 169$&$   824\pm 193$&$   786\pm1384$&$   788\pm1362$&$   872\pm1418$&--\\
00084772+3326000&$    4608$&$ -0.052\pm  0.077$&$ -0.029\pm  0.074$&$ -0.026\pm  0.080$&$  -648\pm 115$&$  -369\pm  63$&$  -340\pm  63$&$  -554\pm 821$&$  -302\pm 782$&$  -275\pm 847$&--\\
00090421+1055081&$    6310$&$ -0.057\pm  0.077$&$ -0.058\pm  0.072$&$ -0.082\pm  0.079$&$  -871\pm 154$&$  -891\pm 148$&$ -1297\pm 236$&$  -828\pm1122$&$  -846\pm1053$&$ -1191\pm1142$&--\\
00092866+4721208&$    4884$&$  0.003\pm  0.107$&$ -0.004\pm  0.108$&$  0.002\pm  0.111$&$    50\pm  12$&$   -35\pm   9$&$    40\pm  10$&$    36\pm1206$&$   -48\pm1211$&$    27\pm1245$&--\\
00093281+4808068&$    4976$&$ -0.163\pm  0.103$&$ -0.137\pm  0.100$&$ -0.155\pm  0.105$&$ -2241\pm 532$&$ -1828\pm 421$&$ -2111\pm 510$&$ -1866\pm1179$&$ -1573\pm1149$&$ -1776\pm1205$&--\\
00102637+2859164&$    7507$&$ -0.097\pm  0.076$&$ -0.099\pm  0.073$&$ -0.133\pm  0.079$&$ -1843\pm 322$&$ -1893\pm 318$&$ -2662\pm 484$&$ -1668\pm1322$&$ -1710\pm1257$&$ -2299\pm1362$&--\\
00103277+2859464&$    6677$&$ -0.014\pm  0.103$&$  0.003\pm  0.100$&$ -0.018\pm  0.104$&$  -199\pm  47$&$    63\pm  14$&$  -267\pm  64$&$  -218\pm1579$&$    42\pm1539$&$  -284\pm1600$&--\\
00111259-3334428&$    7586$&$ -0.042\pm  0.107$&$ -0.038\pm  0.104$&$ -0.019\pm  0.107$&$  -763\pm 188$&$  -683\pm 164$&$  -329\pm  81$&$  -737\pm1873$&$  -666\pm1811$&$  -334\pm1876$&--\\
00121573+2219187&$    7258$&$ -0.063\pm  0.078$&$ -0.055\pm  0.072$&$ -0.056\pm  0.079$&$ -1121\pm 201$&$  -962\pm 159$&$  -993\pm 181$&$ -1051\pm1302$&$  -912\pm1207$&$  -939\pm1322$&--\\
00143187-0044156&$    3589$&$ -0.097\pm  0.109$&$ -0.101\pm  0.109$&$ -0.106\pm  0.112$&$  -893\pm 224$&$  -935\pm 235$&$  -987\pm 254$&$  -798\pm 897$&$  -832\pm 902$&$  -873\pm 926$&--\\
00144010+1834551&$    5037$&$ -0.056\pm  0.065$&$ -0.051\pm  0.057$&$ -0.058\pm  0.066$&$  -703\pm 105$&$  -638\pm  84$&$  -726\pm 110$&$  -646\pm 753$&$  -589\pm 662$&$  -667\pm 768$&--\\
00155127+1605232&$    3838$&$  0.108\pm  0.101$&$  0.112\pm  0.102$&$  0.106\pm  0.105$&$   830\pm 193$&$   857\pm 201$&$   816\pm 197$&$   954\pm 895$&$   989\pm 901$&$   936\pm 927$&--\\
00164417+0704337&$    3598$&$  0.019\pm  0.084$&$  0.021\pm  0.082$&$  0.016\pm  0.087$&$   133\pm  26$&$   154\pm  29$&$   108\pm  22$&$   155\pm 698$&$   177\pm 678$&$   130\pm 719$&--\\
00165087-0516060&$    3597$&$  0.159\pm  0.132$&$  0.159\pm  0.134$&$  0.161\pm  0.135$&$  1088\pm 328$&$  1088\pm 336$&$  1099\pm 342$&$  1316\pm1089$&$  1315\pm1112$&$  1332\pm1120$&--\\
00180131-5905008&$    8707$&$ -0.011\pm  0.090$&$ -0.021\pm  0.086$&$ -0.050\pm  0.091$&$  -185\pm  38$&$  -401\pm  79$&$ -1022\pm 214$&$  -219\pm1804$&$  -427\pm1730$&$ -1002\pm1828$&--\\
00181053+1817323&$    5150$&$ -0.094\pm  0.089$&$ -0.063\pm  0.088$&$ -0.080\pm  0.092$&$ -1224\pm 251$&$  -790\pm 160$&$ -1020\pm 218$&$ -1110\pm1057$&$  -747\pm1049$&$  -943\pm1097$&--\\
00183335-0616195&$    6064$&$  0.119\pm  0.121$&$  0.116\pm  0.122$&$  0.099\pm  0.124$&$  1457\pm 406$&$  1427\pm 401$&$  1235\pm 353$&$  1663\pm1684$&$  1624\pm1705$&$  1378\pm1736$&--\\

\hline
\multicolumn{11}{l}{Table~\ref{tab:distance} is available in its entirety online. A portion is shown here for guidance regarding form and content.}
\end{tabular}}
\end{table*}

\begin{figure*}
\centering
\includegraphics[width=0.9\columnwidth, angle=-90]{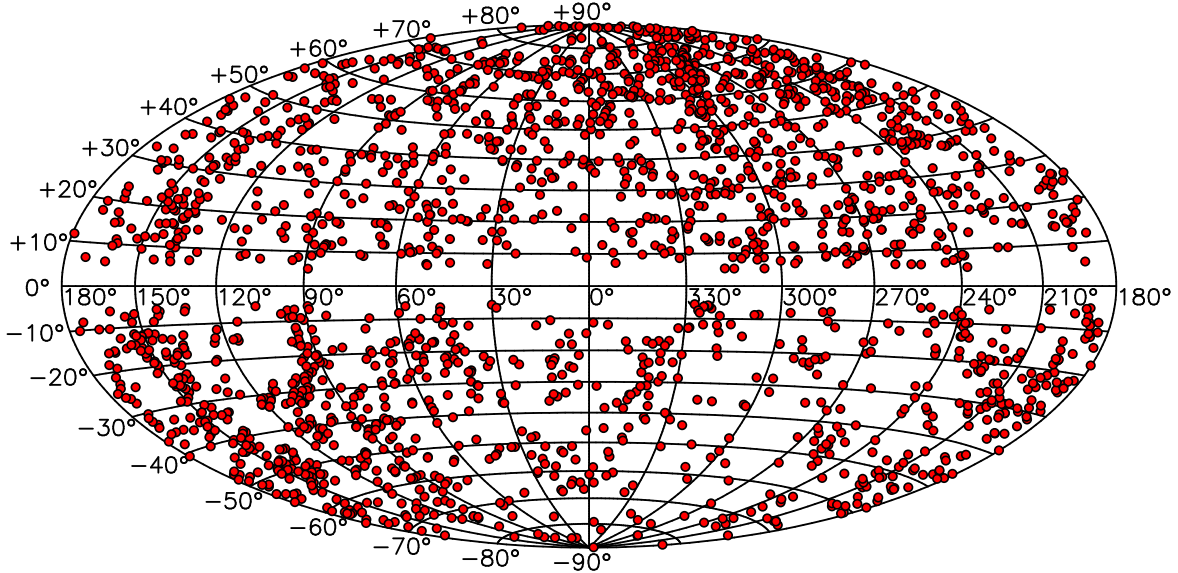}
\caption{Sky coverage of the 2,062 2MTF galaxies, in Galactic
  coordinates using an Aitoff projection.}
\label{fig:sky}
\end{figure*}

\begin{figure}
\centering
\includegraphics[width=0.8\columnwidth, angle=-90]{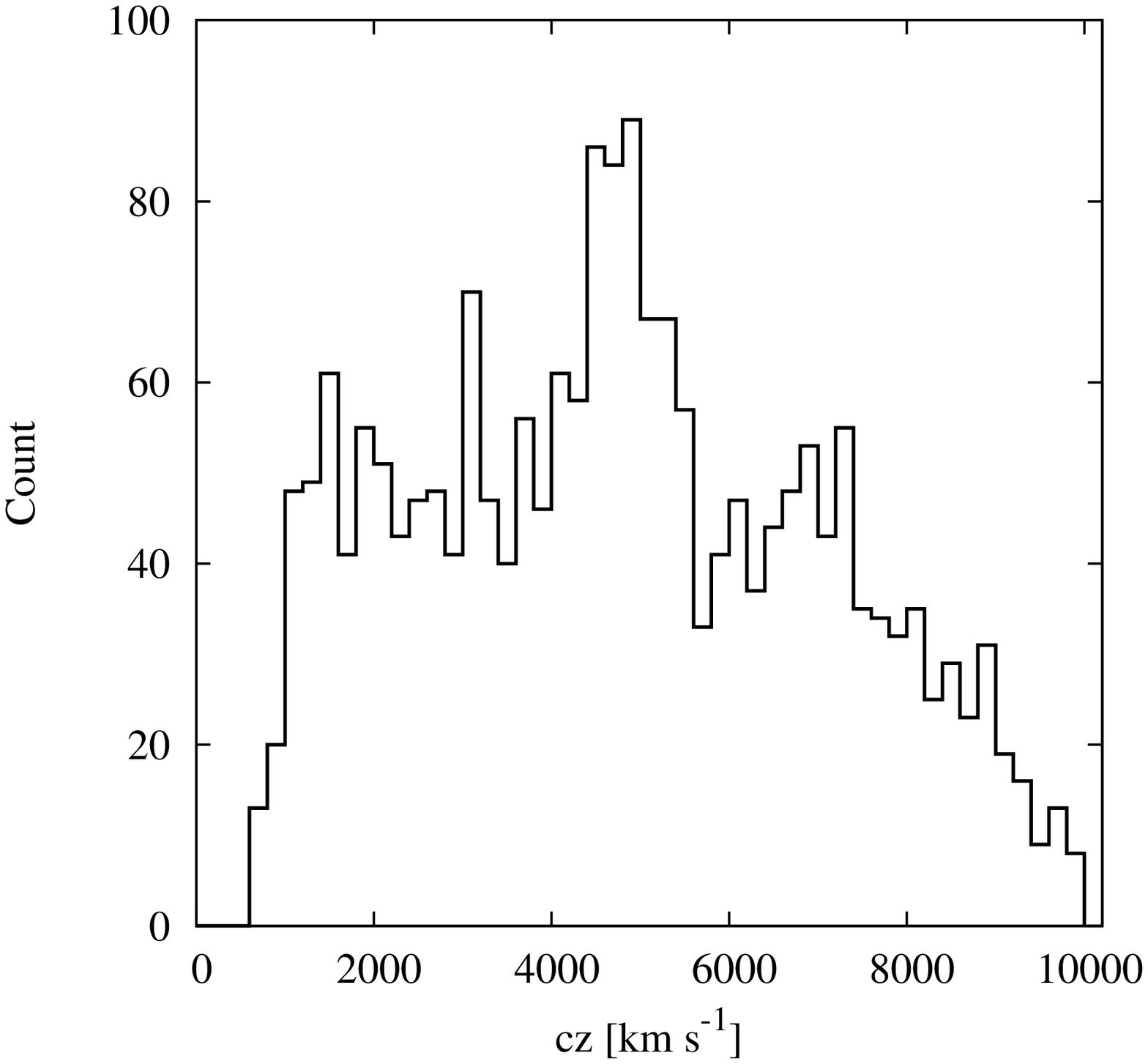}
\caption{The redshift distribution of 2,062 2MTF galaxies in the CMB frame.}
\label{fig:redshift}
\end{figure}

Fig.~\ref{fig:sky} and Fig.~\ref{fig:redshift} show the sky coverage and redshift 
distribution, respectively, of the final 2MTF catalogue which contains 
2,062 galaxies with good data quality. As shown in Fig.~\ref{fig:sky}, the 2MTF catalog provides a fairly 
uniform distribution across the sky with a narrow gap near the Galactic plane ($|b| \leq 5^\circ$).
In the southern sky $\delta < -40^\circ$, the target density is lower because we have only Parkes 
observations in this area. Galaxies in the final 2MTF catalogue cover a CMB velocity range 
of 600 \kms$< cz < 10,000$ \kms\ with a mean velocity of $cz = 4,805$ \kms.

\begin{figure}
\centering
\includegraphics[width=0.8\columnwidth, angle=-90]{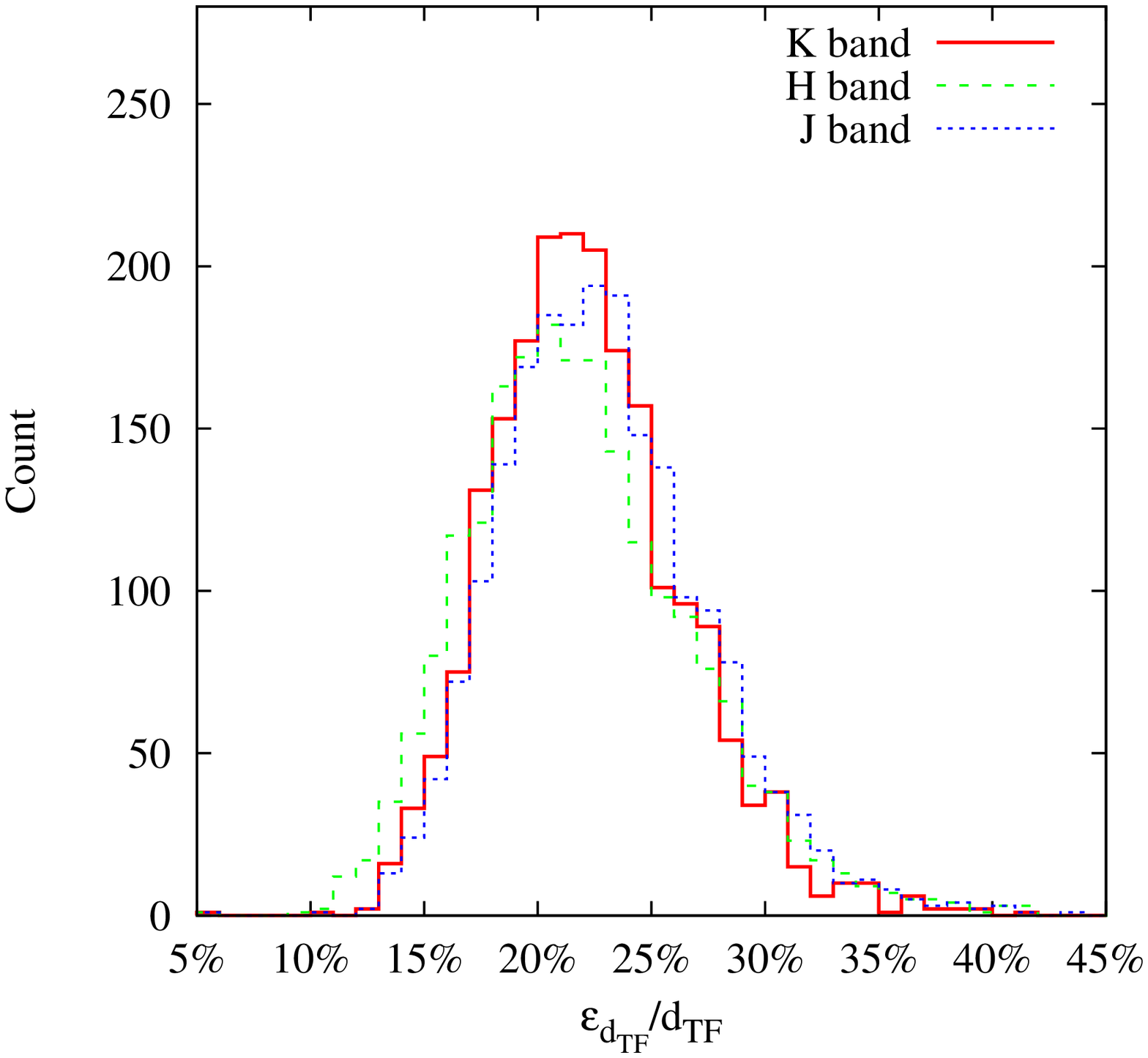}
\caption{Relative errors of the TF distances measured in the $K$ (red solid line), $H$ (green dashed line) and $J$ (blue dotted line) 
bands.}
\label{fig:relative}
\end{figure}

Fig.~\ref{fig:relative} shows the 
relative errors of the linear TF distances measured in the three 2MASS bands. The mean relative errors are 22\%, 22\% and 
23\% in the $K$, $H$ and $J$ bands respectively.

\section{Comparison with Cosmicflows-3}
\label{sec:compare}
We compare the 2MTF redshift-independent 
distances with the distances listed in the Cosmicflows-3 catalogue \citep{Tully2016}. The Cosmicflows-3 catalogue includes around 
17,700 distances, measured by various authors and various methods, mainly Tully-Fisher 
and Fundamental Plane. 

\begin{figure*}
\centering
\includegraphics[width=0.6\columnwidth, angle=-90]{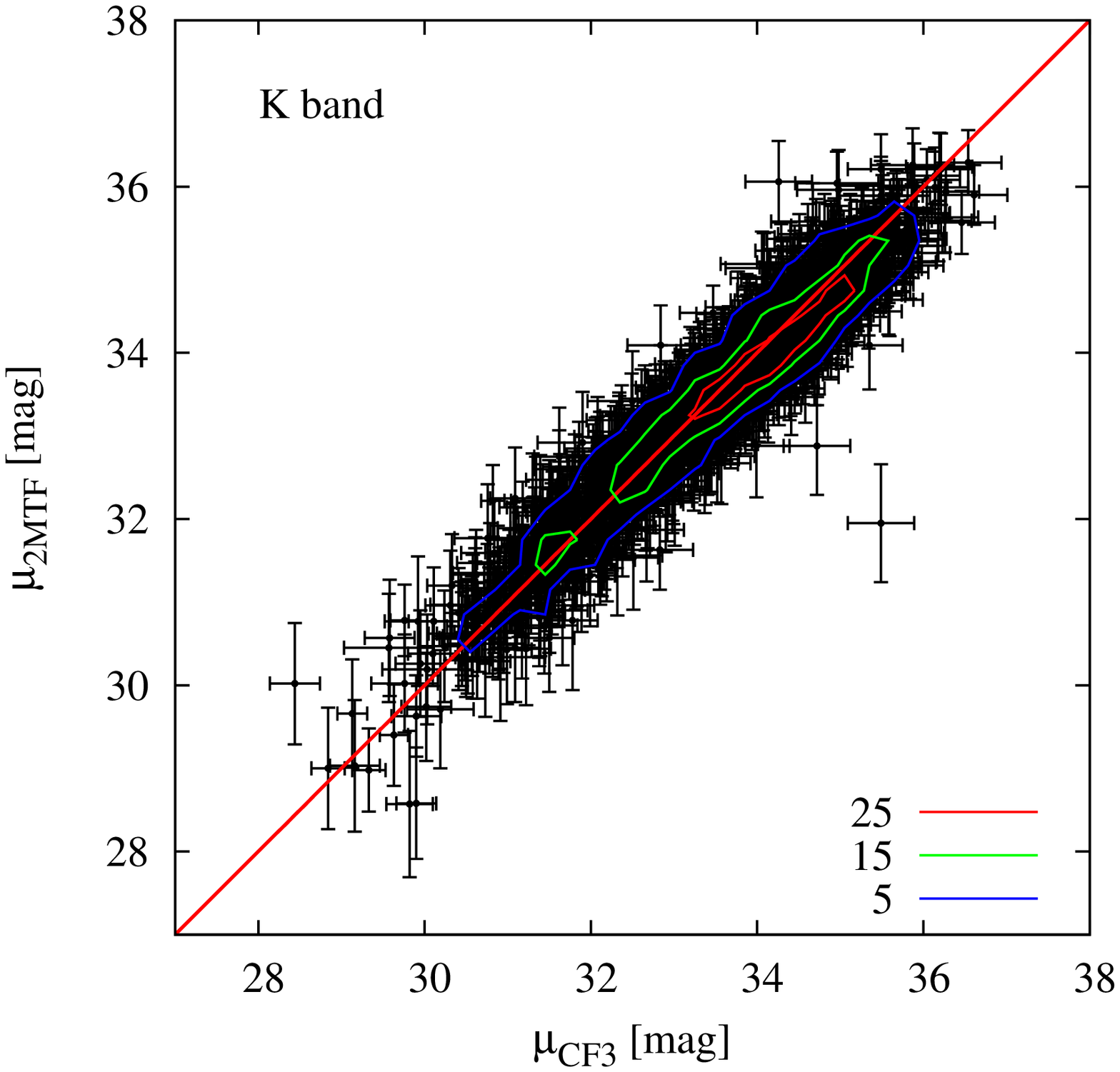}
\includegraphics[width=0.6\columnwidth, angle=-90]{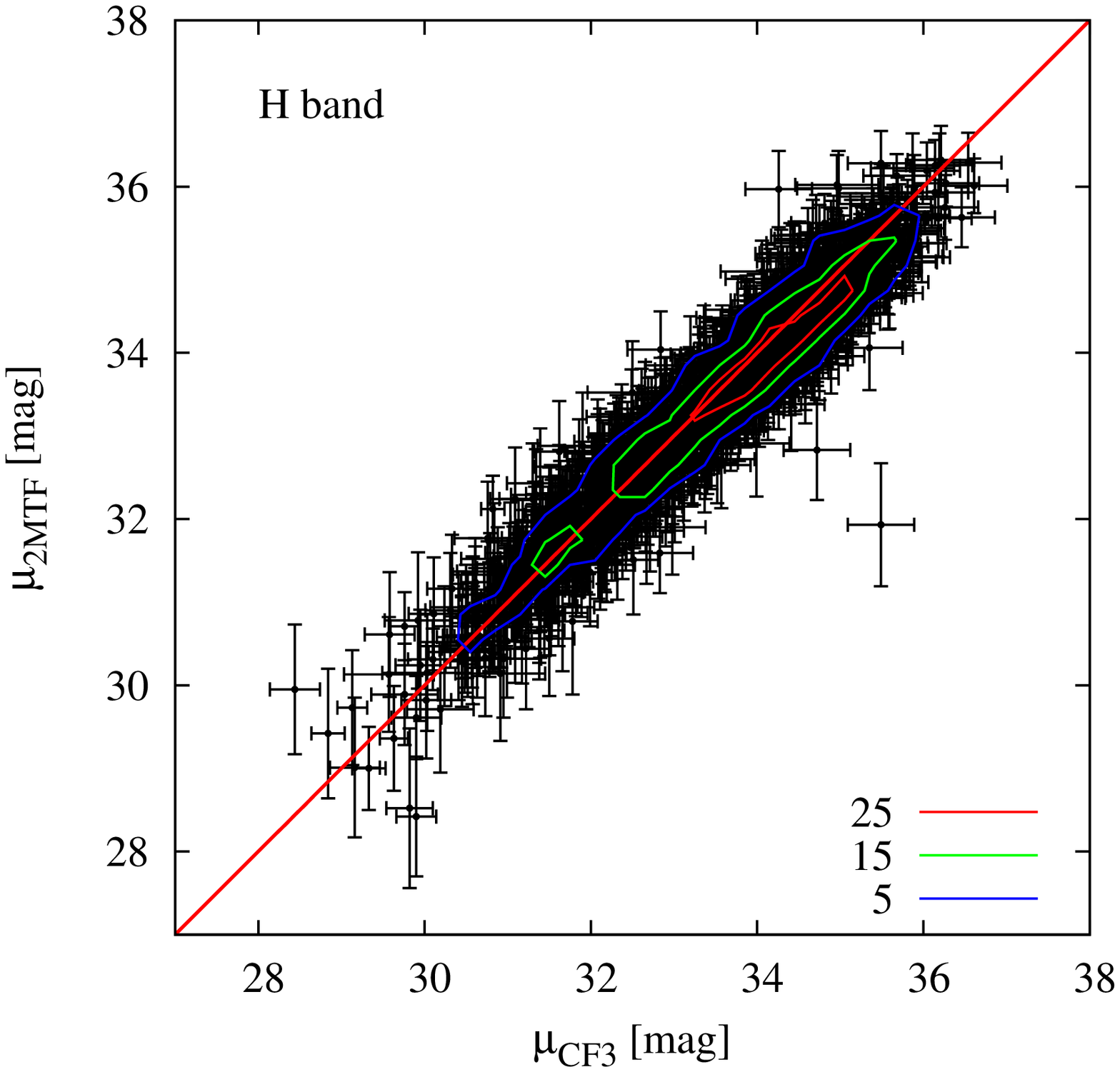}
\includegraphics[width=0.6\columnwidth, angle=-90]{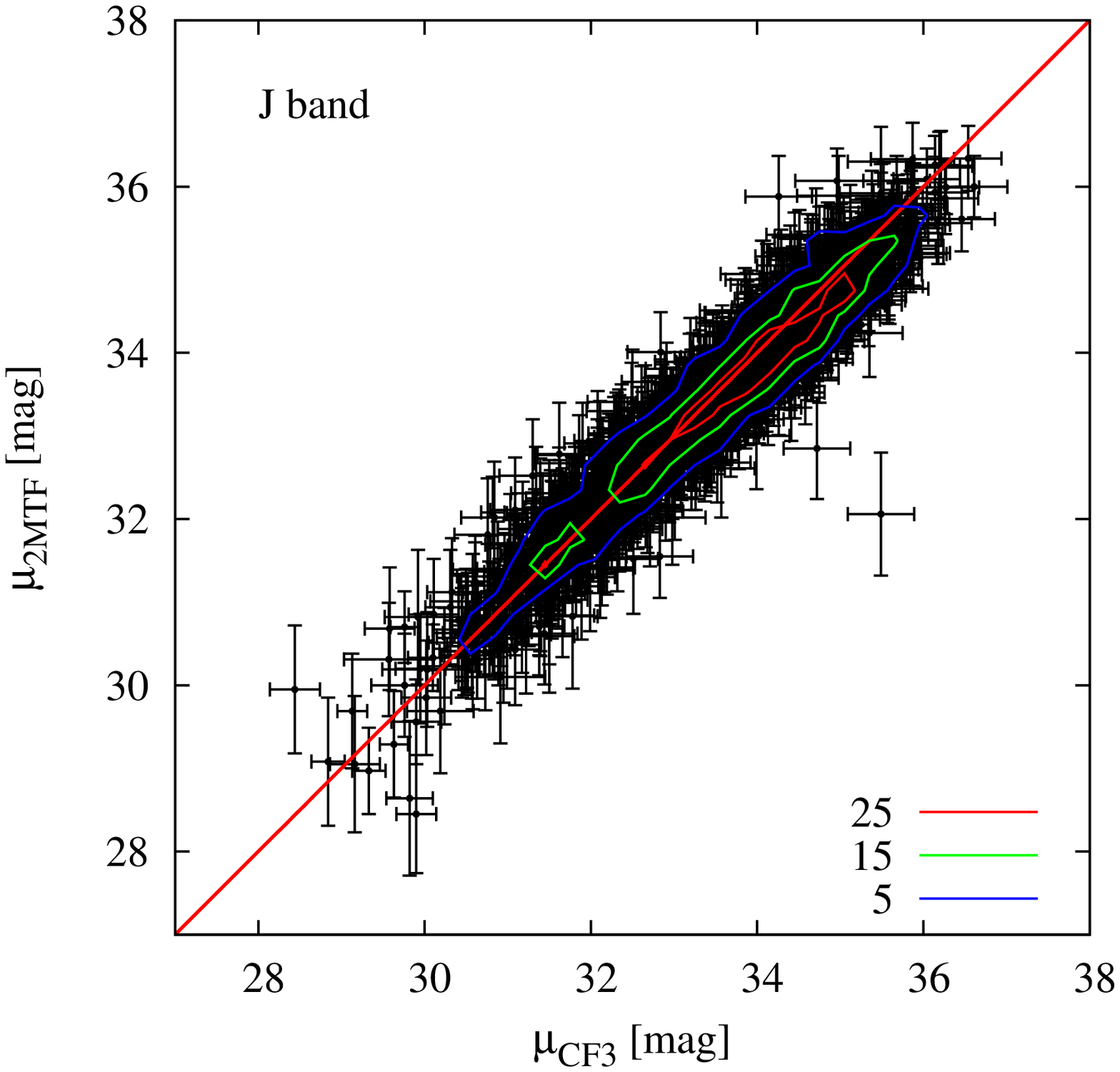}
\caption{Comparison of the distance moduli for 1,117 galaxies common between Cosmicflows-3 and 2MTF in 
the $K$, $H$ and $J$ bands, respectively. The red solid line indicates equality. The galaxy number density in 0.5 mag $\times$ 0.5 mag cells is represented by 
the colour contours. The outlier far away from the equality line at the bottom right is the galaxy 2MASX J00383973+1724113 
for which we give the flag `W' in Table~\ref{tab:distance}.}
\label{fig:CF3}
\end{figure*}

Cross-matching 2MTF with Cosmicflows-3 by sky position and redshift, we found 1,117 common galaxies within a velocity difference $|cz| \leq 150$ \kms. A comparison of the distance moduli of the 2MTF and Cosmicflows-3 measurements reveals only a few outliers far from the line of distance equality. On the whole, there is excellent agreement as shown in Fig.~\ref{fig:CF3}. With a 3$\sigma$ outlier clip, the mean difference between Cosmicflows-3 and 2MTF 
measurements are $0.03 \pm 0.01 $ mag, $0.06 \pm 0.01$ mag and  $0.06 \pm 0.01$ mag in the $K$, $H$ 
and $J$ bands respectively. The corresponding scatter is 0.35, 0.34 and 0.33 mag, respectively. For linear distances, We find the Cosmicflows-3 measurements 
are 1.4\%, 2.8\% and 2.8\% larger than 2MTF distances in three bands respectively. These results are consistent with those 
reported by \citet{Qin2019}.

\begin{figure}
\centering
\includegraphics[width=0.8\columnwidth, angle=-90]{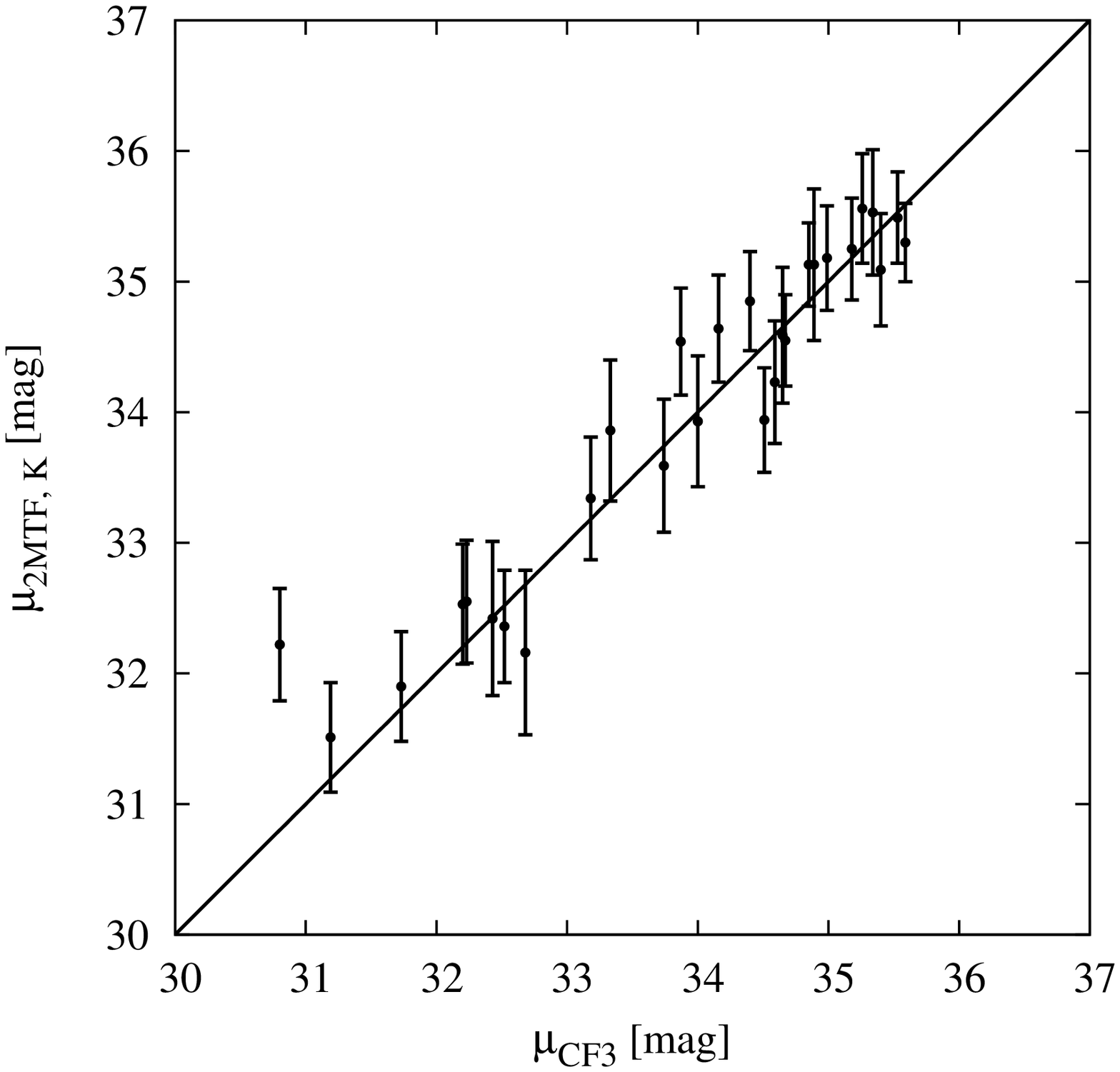}
\caption{Comparison of 28 distances moduli for galaxies measured in 2MTF ($K$ band) 
which host SNIa catalogued in Cosmicflows-3. The solid line indicates equality.}
\label{fig:SNIa}
\end{figure}

Cosmicflows-3 also contains a compilation of 391 supernova type Ia (SNIa) distances for redshifts 
$z \leq 0.1$, with the distance modulus provided being the average of multiple measurements, where available.
Modern SNIa distances can be highly accurate \citep[e.g.][]{Jha2007,Amanullah2010}, and thus allow a more stringent test of 2MTF data quality.
A cross-match between 2MTF and Cosmicflows-3 SNIa distances finds 28 galaxies which have hosted SNIa events. We 
present the comparison of the distance moduli in Fig.~\ref{fig:SNIa}. 
Again, no significant systematic bias in 2MTF distance measurements is found, with the mean difference between 
2MTF and Cosmicflows-3 measurements being $0.12 \pm 0.08 $ mag with a standard deviation of 0.40 mag. However, the outlier in this plot
is 2MASS 09220265+5058353 which hosted SN 1999b. According to the NASA/IPAC 
Extragalactic Database (NED), the distance moduli measured using SN 1999b vary from 30.20 to 31.16 mag, compared with our larger $K$-band 2MTF measurement of $32.22 \pm 0.43$ mag. Neglecting this 
outlier, the mean difference between 2MTF and Cosmicflows-3 SNIa measurements is only $0.08 \pm 0.06$ mag with a 
standard deviation of 0.32 mag. In linear distance space, we find the Cosmicflows-3 SNIa measurements are 
3.7\% smaller than 2MTF distances in mean, the opposite trend of the whole Cosmicflows-3 sample. However, the difference 
is not significant.

\section{Bulk flow in the full 2MTF sample}
\label{sec:bulk}
Bulk flow is measured from the dipole of the peculiar velocity field, and represents the averaged peculiar motion of the 
galaxies in the survey volume with respect to the CMB. In standard theory, it is believed to arise from acceleration induced by  mass distributions within and beyond the sample volume, and therefore provides a useful probe of the mass distribution in the local Universe. It is also 
a powerful tool to test different cosmological models \citep[e.g.][and references therein]
{Hudson2004, Ma2013, Ma2014, Hong2014, Scrimgeour2016, Qin2018}. The high-accuracy distance estimates, the well-defined selection function, and the uniform sky coverage of 2MTF makes this survey easier to use for bulk flow measurements. 

Using the preliminary 2MTF sample, \citet{Hong2014} measured the bulk flow at the depths of 
20$\hMpc$, 30$\hMpc$ and 40$\hMpc$, and found the amplitude to be within the range of the $\Lambda$CDM 
model. \citet{Qin2018} combined the 2MTF catalog with the 6dFGSv survey, making 
an accurate bulk flow measurement at the scale of 40$\hMpc$, and again showed consistency with the 
$\Lambda$CDM model.

In this paper, we measured the bulk flow of the final 2MTF sample using the $\chi^2$ minimization method of \citet{Hong2014}: 
\begin{equation}
\label{eq:chi2}
\chi^2 = \sum_{i=1}^{N}\frac{\left[\log(d_{z,i} / d_{\mathrm{model}, i})- \log(d_{z,i} /
    d_{\mathrm{TF}, i})\right]^2 \cdot w^r_i w^d_i} {\sigma_i^2 \cdot \sum_{i=1}^{N}(w^r_i w^d_i)},
\end{equation}
where $d_{\mathrm{model}, i}$ is the model-predicted distance of the $i$'th galaxy, $\sigma_i$ is the error of $i$'th galaxy's 
logarithmic distance ratio $\log(d_{z}/d_{TF})$, $w^r_i$ is the weight to make the sample's 
weighted redshift distribution match a Gaussian function at a number of depths (here we choose $R_I =$ 20, 30 and 40$\hMpc$), $w^d_i$ is the weight to correct for the effect of the slightly different 
number density in the northern and southern sky areas \citep[see Section 4.1 of][for more details]{Hong2014}.
Instead of using the peculiar velocity data from individual bands, we combine the data from all $K$, $H$ and $J$ bands 
into a `3-bands combined sample' for greatest accuracy. In the combined sample, every galaxy is used 
three times with different peculiar velocities measured from three different bands, with each peculiar velocity counted separately during the minimization procedure. As the 
errors in our TF distances are log-normal, this fit process works in log space instead of linear space. The fitting method is less sophisticated than the $\eta$MLE technique of \citet{Qin2018} which can, in principle, model generic non-normal error distributions. But, since the major errors in 2MTF are the multiplicative distance errors, the current technique is more than adequate. Moreover, it allows a more straightforward comparison with our previous results. 
The errors of the bulk flow velocities were estimated using the jackknife method with 50 sub-samples, randomly selected 
by removing 2\% of the 2MTF sample each time.

The theoretical prediction of the bulk flow amplitude varies with the depth of the galaxy sample \citep{Ma2014}, 
so our theoretical curve was calculated using a Gaussian window function $W(kR) = \exp(-k^2R^2)/2$ 
and a matter power spectrum $P(k)$ generated by the CAMB package \citep{Lewis2000}: 
\begin{equation}
v_{\mathrm{rms}}^2=\dfrac{H_0^2 f^2}{2\pi^2}\int W^2(kR)P(k)\textrm{d}k,
\end{equation}
where $k$ is the wavenumber, $H_0$ is the Hubble constant and $f=\Omega_\mathrm{m}^{0.55}$ is the linear
growth rate \citep{Li2012, Hong2014}. For the calculation of the theoretical curves, we adopt the cosmological 
parameters reported by \citet{Planck2016} with $\Omega_{m}=0.308$, $\Omega_{b}=0.0484$, 
$n_s=0.9677$ and $\rm{H}_0 = 67.81~\rm{km}~\rm{s}^{-1}~\rm{Mpc}^{-1}$.

\begin{figure}
\centering
\includegraphics[width=0.65\columnwidth, angle=-90]{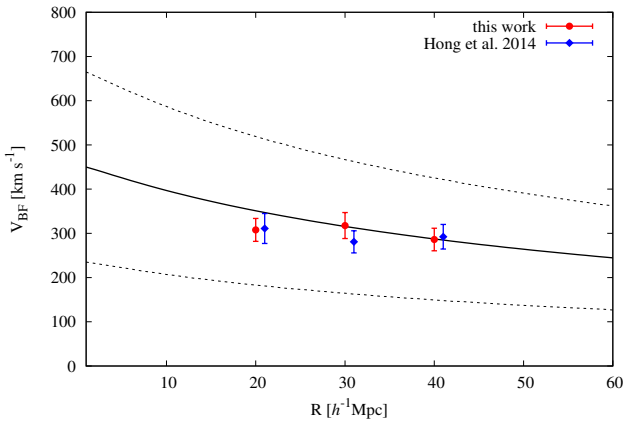}
\caption{The bulk flow amplitude measured using the three-band combined sample at the scale of 20, 30 and $40\hMpc$, 
respectively (red dots). The results of \citet{Hong2014} using the preliminary 2MTF sample are plotted using the blue diamonds 
(shifted $1\hMpc$ right for clarity). The solid line shows the $\Lambda$CDM theoretical prediction, and the dashed lines are 
the 68\% confidence level of the sample expectation.}
\label{fig:bulk}
\end{figure}

The best-fit bulk flow velocities for the three different depths are presented in the Table~\ref{tab:bulk} together with the results measured by \citet{Hong2014} 
using the preliminary 2MTF sample. We also plot the bulk flow velocity amplitudes in Fig.~\ref{fig:bulk}, where the solid line 
is the $\Lambda$CDM prediction and the dashed lines indicate the $\pm 1\sigma$ points of the theoretical line (i.e. $\pm 34$ per cent).
The best-fit bulk flows agree with our previous measurements but have smaller uncertainties, mainly because the new sample 
has a larger size and the updated ALFALFA data has higher accuracy than the previous archival data. Again, our results 
are consistent with the $\Lambda$CDM model prediction at 68\% confidence level, which supports the correctness 
of the $\Lambda$CDM model in the local Universe.

\begin{table}
\caption[]{Bulk flow amplitudes measured in the 2MTF sample.}
\label{tab:bulk}
\centering
\begin{tabular}{ccc}
\hline
\hline
 & \citet{Hong2014} & this work\\
 & [\kms] & [\kms]\\
\hline
$R_I=20~\hMpc$ & $310.9 \pm 33.9$ & $307.8 \pm 25.9$ \\
$R_I=30~\hMpc$ & $280.8 \pm 25.0$ & $317.6 \pm 29.4$ \\
$R_I=40~\hMpc$ & $292.3 \pm 27.8$ & $286.1 \pm 25.5$ \\
\hline
\end{tabular}
\end{table}

\section{Summary}
\label{sec:conc}
2MTF is an all-sky survey which provides accurate Tully-Fisher distances for galaxies in the local Universe with a well-defined selection function. Due to the use of near-infrared photometry, the survey has more uniform sky coverage than similar previous surveys. The Zone of Avoidance around the Galactic plane is only $|b| \leq 5^\circ$. 

Together with good quality \HI data from new observations, the ALFALFA survey 
and high signal-to-noise ratio archival data, 2MTF provides high-accuracy distance measurements for 2,062 nearby spiral galaxies. The 
mean relative error in the final 2MTF sample is 22\%.
We compare our measurements with the published distances in Cosmicflows-3, which represents the largest compendium of nearby galaxy distances. We find no substantial systematic difference between 2MTF and Cosmicflows-3 distances. Higher accuracy SNe Ia distances were also compared for 28 cross-matched objects. Again, no substantial systematic difference 
was found.

The best-fit bulk flow velocity amplitudes are  
$V = 308 \pm 26$ \kms, $V = 318 \pm 29$ \kms,
and $V = 286 \pm 25$ \kms at depths of $R_I = $ 20, 30 and 40$\hMpc$ respectively, consistent with our previous measurements using the preliminary 2MTF sample but with higher accuracy. The 
fit results agree with the $\Lambda$CDM prediction at the 68\% confidence level.\\

\section*{Acknowledgments}
The authors wish to acknowledge the contributions of John Huchra (1948 - 2010) 
to this work. The 2MTF survey was initiated while KLM was a post-doc working 
with John at Harvard, and its design owes much to his advice and insight. This 
work was partially supported by NSF grant AST- 0406906 to PI John Huchra.

Parts of this research were conducted by the Australian Research
Council Centre of Excellence for All-sky Astrophysics (CAASTRO),
through project number CE110001020, and the Australian Research
Council Centre of Excellence for All Sky Astrophysics in 3 Dimensions (ASTRO 3D),
through project number CE170100013. TH was supported by the National
Natural Science Foundation of China (Grant No. 11473034, U1731127), the Key
Research Program of the Chinese Academy of Sciences (Grant No.
QYZDJ-SSW-SLH021), the strategic Priority Research Program
of Chinese Academy of Sciences (Grant No. XDB23010200),
and the Open Project Program of the Key Laboratory of FAST,
NAOC, Chinese Academy of Sciences.

\bibliographystyle{apj}
\bibliography{bibfile}

\begin{appendix}

\end{appendix}

\label{lastpage}
\end{document}